\documentclass[traditabstract,oldversion]{aa} 

\usepackage{graphicx}

\usepackage{txfonts}
\def\bibl{\parindent=0pt \hangindent=0.150 in }
\begin{document}

\title{Stripping a debris disk by close stellar encounters in an open stellar cluster }

\author{Jean-Fran\c cois Lestrade 
       \inst{1},
         Etienne Morey \inst{1}, 
        Antoine  Lassus \inst{2},
     \and
        Naron Phou   \inst{2}
} 

\offprints{J.-F. Lestrade, e-mail : jean-francois.lestrade@obspm.fr}

\institute{Observatoire de Paris/LERMA - CNRS, 61 av. de l'Observatoire, F75014, Paris, France\\
\email{jean-francois.lestrade@obspm.fr}
\and 
UPMC, Universit\'e Pierre et Marie Curie, 4 place Jussieu,  F75005, Paris, France\\
}

\date{Received April 2010 ; accepted June 2011}

\abstract{A debris disk is a constituent of any planetary system surrounding a main sequence star. 
We study  whether close stellar encounters  can disrupt and  strip a debris disk of its planetesimals 
in the expanding open cluster of its birth with a decreasing star number density over 100~Myrs. 
Such stripping would affect the dust production and hence detectability of the disk. 
We tabulated the fractions of planetesimals stripped off  during stellar flybys of miss distances 
between 100 and 1000 AU and for several  mass ratios of the central to passing stars. 
We then  estimated  the numbers of close stellar encounters over the lifetime of several expanding 
open clusters characterized  by their initial star densities. 
We found that a standard disk, with inner and outer radii of 40 and 100~AU, suffers no loss of planetesimals 
over 100~Myrs around a star born in a common embedded cluster with star density $\le 1000~pc^{-3}$. 
In contrast, we found that such a disk  is  severely depleted of its planetesimals over this timescale 
around a star born in an Orion-type  cluster where the  star density is $>20~000~pc^{-3}$. 
In this environment, a disk loses  $>97$\% of its planetesimals around an M-dwarf, 
$>63$\%  around a solar-type star, and $>42$\% around an A-dwarf, over 100~Myrs. 
We roughly estimate that two-thirds of the stars
may be born in such high star density clusters. This might explain in part 
why fewer debris disks are observed around lower mass stars.

}

\titlerunning{Stripping a debris disk}
\authorrunning{Lestrade et al.}

\maketitle


\section{Introduction}

A debris disk surrounding a main sequence star is the collection of planetesimals 
(comets or asteroids) that are the leftovers from an early phase of planet formation.
A debris disk is a constituent  of any planetary system in the core-accretion theory of  planet formation. 
It is the  analogue of  the Kuiper Belt   beyond Neptune, or of the asteroid belt 
between Mars and Jupiter, in our Solar System. The formation and evolution 
of debris disks and planetary systems are inter-related in theory, and can provide complementary 
insight into each other.

\begin{figure*}[t]
\centering
\includegraphics[width=17.cm, angle=0, bb=0 300 595 842]{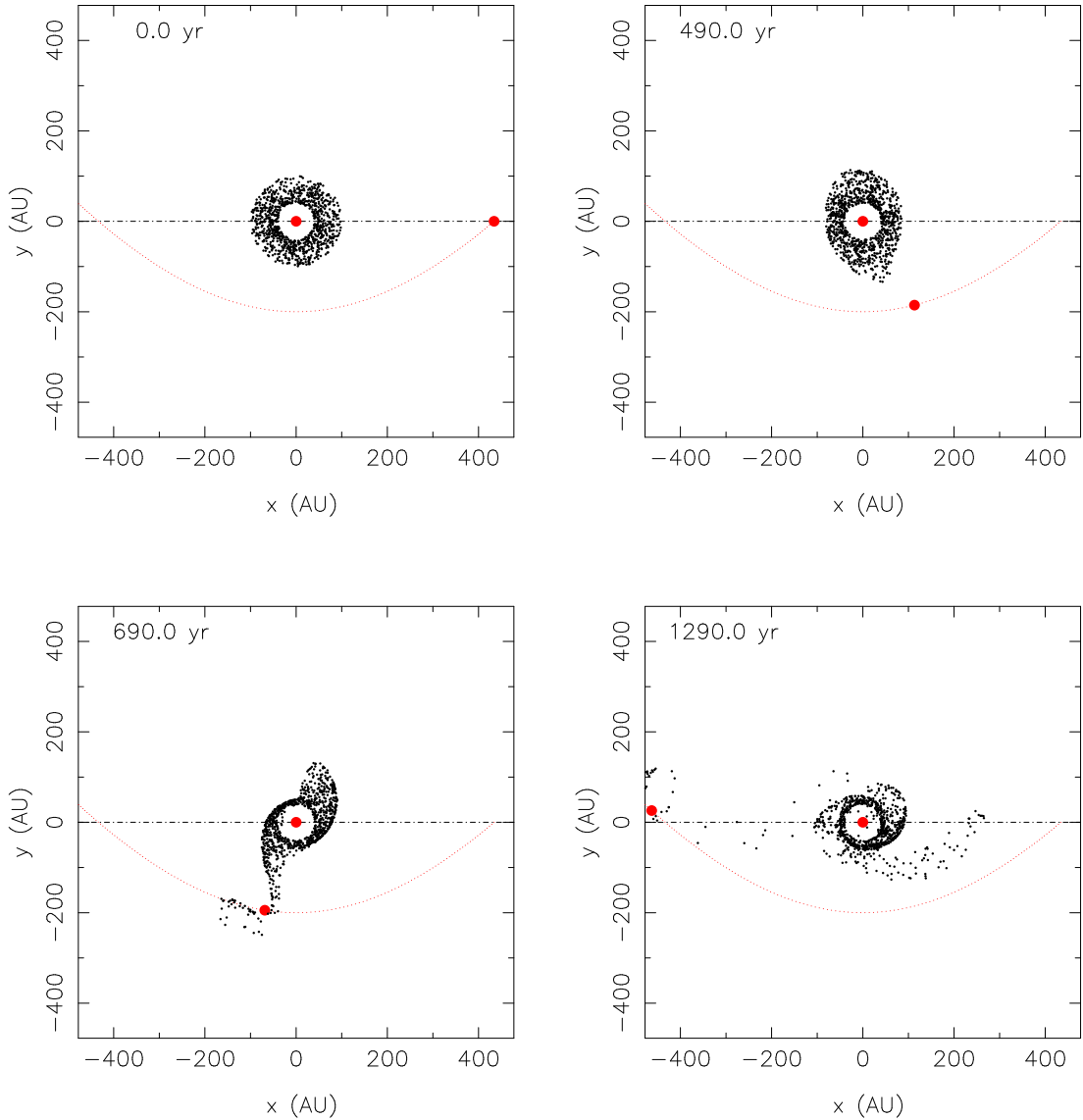}
\caption{Stellar flyby. The debris disk is non-self-gravitating, and has inner and outer radii of 40 and 100 AU at the start of
the simulation. 
The central star is at the origin of the coordinate system. The trajectory of the passing star is parabolic, 
coplanar and  prograde with respect to the disk.  The mass ratio of the two stars is unity. 
 At the closest approach, the miss distance  is 200 AU and the resulting maximum velocity is 4.2 km/s. 
The fraction of planetesimals stripped  off the disk and lost by the system is 13\% during this close stellar encounter. 
\label{flyby}} 
\end{figure*}

The dust content of a debris disk is the key to its observability. 
Small dust grains scatter the light of the central star and  make the disk visible at optical wavelengths. 
Large grains are  heated by the central star and efficiently re-radiate in the 
far-IR and (sub)millimeter domains where they are observed. It is thought that dust must be   
continually or episodicly  replenished by  mutual collisions between planetesimals because dust grains are
removed on short timescales (a few millions years or less).  
Disks have to be several-fold dustier than the Kuiper belt $(\times 100$)
to be detectable with our  instrumental sensitivities, even for the nearest stars. 
Nonetheless, bright  disks are detected  around $32\pm5$\% of A-type dwarfs (Su et al 2006, Wyatt et al 2003),  
$16\pm2.8$\% of solar-type dwarfs (Bryden et al 2006,  Trilling et al 2008, Najita \& Williams 2005)
and $\le$5\% of M-type dwarfs (Lestrade et al 2006 and 2009, Gautier et al 2007) according to searches 
for cold dust at  $\lambda~=~70\mu$m by Spitzer
and at (sub)millimeter wavelengths by radiotelescopes.  
Debris disks were recently reviewed by  Wyatt (2008). 
 
 The mechanisms responsible for  enhanced collisional activity  and  for grinding planetesimals to dust  in debris disks 
 remain unclear. The statistics of disk detections just recalled cannot be accurately interpreted  unless
 these mechanisms are identified. It has been proposed that unseen planets dynamically stir the disk and  
 generate copious amount of dust that evolve under gravity, radiation pressure, P-R, and stellar-wind drags
 (Moro-Mart\'in \& Malhotra, 2002;  Wyatt  2003; Mustill \& Wyatt 2009; Kennedy \& Wyatt 2010). Steady-state  equilibrium between 
 collisional cascades of planetesimals  and dust removal
 processes  can explain how   disks fade with age  (Dominik \& Decin 2003). 
 Orbit resonance crossing of giant planets can trigger  
 abrupt showers of planetesimals in the system, severely depleting the disk and limiting its detectability
 (Gomes et al 2005, Morbidelli et al 2005, Tsiganis et al 2005).  However, the distribution of disk fractional
luminosities with star ages is inconsistent with the idea that this mechanism is common among debris disks (Booth et al. 2010). 
Finally, distant icy-planets can successively form in waves outwards in the disk and generate dust rings (Kenyon \& Bromley, 2002).   

In this paper, we study whether stellar flybys  during the first 100 Myrs of the lifetime of a star, 
while it is still in the expanding open cluster of its birth, can significantly deplete a disk of its planetesimals, 
affecting its dust production and therefore  detectability.  Stellar flybys have been invoked to 
explain the high eccentricity orbits of some Kuiper Belt Objects such as  Sedna (Kenyon \& Bromley, 2004),
 the dynamics of planetary systems (Malmberg et al. 2007, Spurzem et al. 2009), and the structures of  debris disks
 (Larwood, 1997; Mouillet et al. 1997;  Kalas et al 2000; Kobayashi \& Ida,  2001). 
 Impacts of stellar flybys during the first few million years have also been studied in the context of the evolution 
of protoplanetary disks and  
 planet formation (Bonnell et al 2001,  Olczack, Pfalzner \& Eckart 2009). 
 
 For our study, we did not resort  to a full N-body integration 
 of an  expanding cluster with stars surrounded by disks of planetesimals. We instead divided the
 problem into parts by first estimating the fraction of planetesimals stripped off a disk by the passage of a star 
 moving on a parabolic trajectory  in Sect.~2, and second by estimating the probability of
 close stellar encounters in an expanding cluster in Sect.~3.
 Finally, we discuss our results in Sect.~4.

\section{Stripping a debris disk by stellar flybys}

In our model, the  central and the passing stars are considered to be point masses, 
named $m_c$ and $m_p$, respectively. The disk is made of $10^4$  
massless planetesimals randomly distributed, radially and azimutally, with 
the  planetesimal surface number density  $\Sigma(a)=\Sigma_0 {(a/a_0)}^{-1.5}$.  
Hence, the disk is not self-gravitating, {\it i.e.} the orbit of each planetesimal is computed  
under the sole gravitational attractions  of $m_c$ and $m_p$ in the restricted three-body
problem. 

In our simulations, we considered two sizes for the disk : 
a standard configuration with inner and outer radii of 40 and 100~AU, 
and a more compact configuration with respective radii 10 and 40~AU, at the start of the computation.
These characteristic radii are based on limited observations at present. 
Protoplanetary transition disks have inner radii ranging from 15 to 73 AU and outer radii from 30 to 135 AU, 
as measured by the SMA (Andrews et al. 2011).  
More mature debris disks around main sequence stars 
have radii determined  by means of either imaging or  the fitting of their SED, 
and  comprised between $\sim$ 10 AU  and  $\sim$ 300 AU 
({\it e.g.} Dent et al. 2000, Greaves et al. 2005, Liseau et al. 2008, Corder et al. 2009, Marshall et al. 2011).
In addition, disk size may depend on the central star mass; for example, the disk surrounding the M-star AU~Mic
is about three times smaller than the disk around 
the  A-star $\beta$~Pic of the same age (12~Myr) (Augereau \& Beust 2006, Augereau et al. 2001).
However, a correlation between  disk size  and star mass has not yet been established 
because of a  lack of sufficient data. Current surveys with Herschel in the far-IR 
(Matthews et al., 2010, Eiroa et al., 2010) will provide new and important constraints.
We  adopted the two disk configurations, standard and compact, in our simulations 
to help us prepare for any revisions of disk sizes in the future.

Our first suite of simulations was conducted in a three-dimensional (3D) coordinate system where the disk is
confined to the $x$-$y$ plane but the parabolic orbit of the passing star can be in an inclined plane.  
The starting position of the passing star is chosen to be at a large distance 
($> 1000$ AU) to ensure that the disk is not perturbed by an impulsive force at the start of the computation.
The initial velocity for the passing star is set so that, at the closest approach,  
it  reaches  a given miss distance while on a parabolic trajectory. The miss distances were  set
to be between 100 AU and 1000 AU,  and the resulting maximum velocities of the passing star 
at closest approaches  were between 3 and 12 km/s. This maximum velocity is related to the miss distance 
because of the parabolic nature of the orbit.

The equation of motion of a massless particle under the gravitational 
attractions of the central and passing stars were written in a coordinate system 
where the central star is at the origin. These equations were integrated 
numerically using the fourth order Runge-Kutta algorithm described in Press  et al (1992). 
The numerical accuracy of  our implementation of this algorithm was tested 
by computing the total energy of a test star, {\it i.e.} one not gravitationally 
interacting with the disk, set on a high eccentricity orbit with a small semi-major axis
to produce the highest acceleration in the system. We set the time step for our integration
so that the total energy of this test star was conserved to an accuracy lower than 0.1 \%. We also extensively 
compared our Runge-Kutta results to the ones that we obtained first with the computationally expensive  
Euler method.

\begin{figure}[h]
\centering
\includegraphics[width=6.5cm, angle=-90]{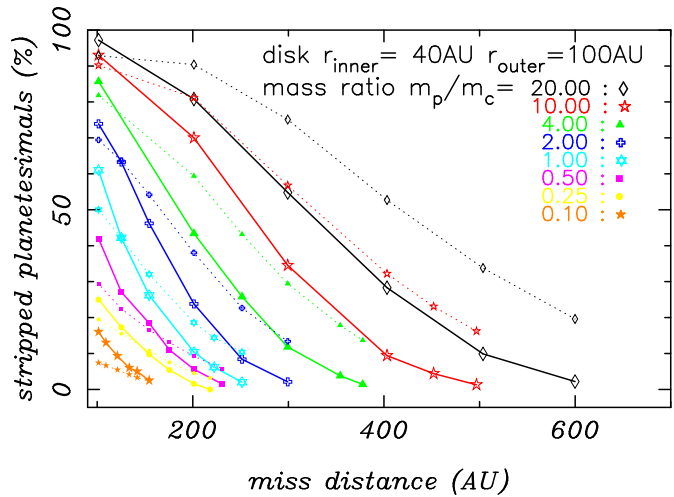}
\caption{Fractions of planetesimals stripped off a disk
during stellar flybys at various miss distances and for various color-coded mass ratios $m_p/m_c$ of the passing to central stars. 
The motions of the disk and of the  passing star are prograde and their orbital planes are coplanar. 
The inner and outer radii of the disk are 40 and 100 AU, respectively, at the start of the simulation. Thick lines connect the points 
computed for a disk with planetesimals initially set on circular orbits. Dotted lines connect the
points computed for a dynamically excited disk with orbits of planetesimals initially set with random eccentricities and inclinations 
(see details at end of Sect.~2).   
\label{strip40_100_dots}}
\end{figure}

\begin{figure}[h]
\centering
\includegraphics[width=6.5cm, angle=-90]{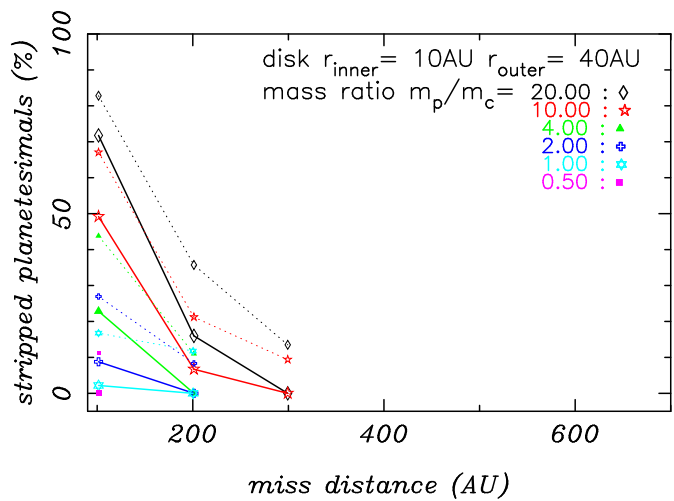}
\caption{Same as Figure~\ref{strip40_100_dots} but for a compact disk with inner and
outer radii of  10 and 40 AU.   
\label{strip10_40_dots}}
\end{figure}

The dynamical evolution of our system is simulated over a few thousand years, long enough for the passing star to reach
the closest approach and move away to a large distance. Our simulation is exemplified in Figure \ref{flyby} for prograde 
and coplanar motions of the disk and passing star. This example shows that the disk is first slowly perturbed and 
then significantly disrupted at closest approach where  the tidal forces 
reach a maximum. Finally the disk settles in to its new configuration with  bound planetesimals on elliptical orbits,
while other planetesimals are either captured by the passing star or ejected on unbound orbits.
We checked that the fractions of stripped planetesimals (captured + ejected) are accurately
 estimated for a disk confined to the $x$-$y$ plane and accurately represented 
by $10^4$ particles in comparing to  simulations with larger numbers of particles.

Our simulations were carried out by varying the miss distance from 100~AU to 1000~AU, 
the sense of circulation between the disk and the passing star (prograde, retrograde), 
the inclination $i$ of the  orbital plane of the passing star ($0^{\circ}$, $30^{\circ}$,
$45^{\circ}$, $90^{\circ}$), and the mass radtio $m_p/m_c$ (20, 10, 4, 2, 1, 0.5, 0.25, 0.1).
We found that all stellar flybys
with retrograde motions are inefficient  leading to very low fractions of stripped planetesimals. This is 
because they do not move in concert with  the passing star and are not gravitationally perturbed 
for a sufficiently long period of time,  as also noted by other authors in similar circumstances 
(Toomre \& Toomre, 1972, Beust \& Dutrey, 2006, Reche et al., 2009).  
All flybys with prograde motion and  inclination $i \ge 45^{\circ}$ are also inefficient.  
Only flybys with prograde 
motion and inclination   $ i < 45^{\circ}$ lead to significant stripping whose magnitude  is 
quantified by Figures \ref{strip40_100_dots} and \ref{strip10_40_dots}. These figures show the fraction of stripped planetesimals 
as a function of the miss distance, for the mass ratios $m_p/m_c$ mentioned above, for the standard and compact disks,
and  for inclination $i < 45^{\circ}$ of the orbital plane of the  passing star. 
Clearly, by comparing Figures \ref{strip40_100_dots} and \ref{strip10_40_dots}, 
a compact disk with inner and outer radii of 10 and 40~AU is much less severely depleted  of
its planetesimals than the more extended one.
 
To check our simulations,  we  duplicated the 
results of Larwood and Kalas (2001) for  the debris disk around $\beta$~Pic 
perturbed by a passing star.  We found excellent agreement with their Table~1 for the numbers of particles 
unbound and captured by the passing star,  for all ratios $m_p/m_c$ and inclinations.  
We  also compared our results with the study of 
stripping effects in the Kuiper Belt when the Solar System was in 
its birth aggregate performed by Adams and Laughlin (2001).  They estimate that  40\% of the Kuiper Belt objets with semi-major
axis between 30 and 70~AU are removed  
from the Solar System  in an encounter with a 1 solar mass star at the miss distance  
$\sim 200~AU \sim 350/\sqrt{\pi}~AU$ by using their  fiducial  cross-section $\sigma \sim (350~AU)^2$. 
This fraction is higher than our prediction of 13\% in Figure \ref{strip40_100_dots}. We ascribe this
difference to their model which also includes the scattering  of high eccentricity planetesimals by Neptune
that is not considered in our model.

\begin{table}[t]
\centering
\caption{Stellar flybys of miss distance 200~AU for two initial disk excitations $^{*}$. \label{frac_changes}}
\begin{tabular}{l|c|c} \hline
\hline\\
 $m_p/m_c$             &     \multicolumn{2}{c}{Fractions of stripped planetesimals}  \\\hline
                       &     no excitation      &   excitation \\\hline
                       &                        &               \\
   4                   &      47 \%             &   63 \%     \\
   2                   &      27 \%             &   42 \%     \\
   1                   &      12 \%             &   21 \%    \\
   0.5                 &      5  \%             &   12 \%    \\
   0.25                &     1.2 \%             &   6.2 \%     \\\hline
\end{tabular}
\begin{flushleft}
~~~~~~~~~~~~~~~~$^{(*)}$ standard size disk. 
\end{flushleft}
\end{table}

 Finally, we  studied how the initial conditions in the disk affect the outcome of a flyby since 
the  disk can be initially excited by the dynamical perturbations of a previous flyby. 
We run a new suite of simulations in which planetesimals were set on eccentric and inclined, 
instead of circular, orbits.
Their eccentricities, inclinations, and semi-major axes $a$ were randomly  
distributed between 0.0 and 0.95,  $\pm 30^{\circ}$, and  40 and 100~AU, respectively, at the start of the simulation. 
The planetesimal surface number density $\Sigma(a)$  was the same power-law  as given above and $10^5$ particles 
were used to represent this three-dimensional (3D) disk. These new simulations have yielded fractions of stripped planetesimals
that are somewhat larger than for a disk with no initial excitation, but not exceedingly larger. These new fractions are the points 
connected by dotted lines in Figures~\ref{strip40_100_dots} and \ref{strip10_40_dots}. 
To appreciate the differences, we also present a subset of these points 
in Table~\ref{frac_changes} for a standard disk, miss distance of 200 AU and mass ratios $m_p/m_c =$ 4, 2, 1, 0.5, 
and 0.25, which are representative  examples.
We estimated that  the final depletions after 100~Myr between the two models can change  by at most a factor 
of two depending on the exact history  of encounters.  
Such a change is not important for the assessment of the impact of depletion on the dust production
of a disk; only a depletion factor greater than ten is meaningful for the disk detectability.
For the remaining analysis, we adopted the fractions given
in Figures \ref{strip40_100_dots} and \ref{strip10_40_dots} for a disk with no initial excitation.

\section{Close stellar encounters in an expanding open cluster}

\subsection{Number of stellar encounters over the cluster lifetime}

We do not resort to an N-body calculation to estimate the frequencies of  encounters 
between stars in an open cluster. We instead use the kinetic theory complemented by gravitational focusing.   
In this framework, the  encounter time $t_{enc}$  can be
estimated analytically by  rederiving  Eqs.~8-122  of Binney \& Tremaine (1987, p. 539), which is valid for two stars 
of identical masses, for the  more general case 
of a central star and a passing star of  masses $m_c$ and  $m_p$ moving in a field 
with uniform star number density $n$. The  corresponding encounter rate at 
miss distance $d < d_{enc}$  is 

$$ {1 \over t_{enc}} = 4 \sqrt{\pi} ~ n ~ \sigma  ~ {d_{enc}}^2 + {{2 \sqrt{\pi}  G (m_c +m_p) ~ n ~ d_{enc}} \over {\sigma}} ~~~~~~~~~(1),$$

\noindent where $\sigma$ is the dispersion of the Maxwellian  distribution 
in the stellar velocities of the cluster, and  $G$ is the gravitational constant.
The first term is the collision rate inferred from the simple kinetic theory of a star moving 
at the mean velocity $4 \sigma/\sqrt{\pi}$ (mean velocity of a Maxwellian  with dispersion $\sigma$). 
This expression  is based on the volume swept by the cross-section $\pi d_{enc}^2$ at this mean velocity during $t_{enc}$.   
The second term is gravitational focusing to account for the true
orbits of the two stars. In convenient units, the  total encounter rate is 

$$ {1 \over t_{enc}}  (yr^{-1}) = \rm 1.9 \times 10^{-8}  \Big({n \over {1000~pc^{-3}}}\Big) ~ \Big({\sigma \over {1~km/s}}\Big) ~ \Big({d_{enc} \over {100~AU}}\Big)^2  + ~~~~~~~~~~~~~~~~~~~~ $$

$$ ~~~~~~~~~ {\rm 8.8 \times 10^{-9}  \Big({{m_c+m_p} \over 1~M_{\odot}}\Big) ~ \Big({n \over 1000~pc^{-3}}\Big) ~ \Big({d_{enc} \over 100~AU}\Big) ~ \Big({1~km/s \over \sigma}\Big)}~~~(2).$$

\noindent For our model, we assume that the star number density $n$ is a linear function of time $t$. 
In our notation, this is 

$$n = n_0 f_p - (n_0 - n_1) f_p (t/t_{cl}) ~~~~~~~~~~~~~~~(3),$$  

\noindent where $f_p$ is the stellar fraction for each spectral type range listed in Table \ref{mass_distr}.
With this linear expression of time, 
the  total star number  density  in the cluster decreases from 
$n_0$ at $t=0$  to   $n_1= 0.1~pc^{-3}$ at $t=t_{cl}$,  where  $t_{cl}$ is the cluster lifetime, 
typically 100~Myr (Lada \& Lada 2003). The final value of 0.1~$pc^{-3}$ is the field star density 
as measured in the solar neighbourhood as well as in relatively old open clusters (Abt, 2009). 
The linear dependence adopted is based on the study of the dynamical evolution of the Pleiades
over 1 Gyr by Converse \& Stahler (2011). They found that the central surface density ($pc^{-2}$) of the 
cluster falls linearly with time at a rate such that the initial density drops by tenfold in 200 Myr, 
and then falls also linearly but at a slower rate until 1~Gyr when the cluster is totally dispersed. They also
show that the core radius is almost constant during the first 125 Myr such that the linear dependence
of the surface number density (pc$^{-2}$) transfers to the volume density (pc$^{-3}$). We also note 
that in a similar way,  Kaib \& Quinn (2008) use this {\it a priori} time dependence for their sudy 
of the formation of the Oort cloud in an open cluster environment. 
 
The number of stellar encounters of miss distances $d~<~d_{enc}$ can be tracked by the phase (cycle)  

$$\phi~(d < d_{enc}) =\int_{~0}^{~t_{cl}}~1/t_{enc}~dt ~~~~~~~~~~(4).$$

\noindent Stellar flybys of miss distances~$<~d_{enc}$,   
mean velocity $4 \sigma/\sqrt{\pi}$, and in the environment of varying star density $n$ of eq.(3),   
occur successively at each integer  value taken by this phase.       

\begin{table}[b]
\centering
\caption{Stellar mass distribution of passing stars \label{mass_distr}}
\begin{tabular}{l|c|r} 
\hline\hline\\
Spectral type $m_p$    & Mass range     &   Fraction $f_p$ \\
                       &  ($M_{\odot}$) &            \\\hline
                       &                &               \\
 M8-M5                 & 0.10 - 0.21    &   43.0 \%     \\
 M4-M0                 & 0.21 - 0.47    &   31.5 \%     \\
 K8-K0                 & 0.47 - 0.80    &   12.4 \%    \\
 G8-G0-F0              & 0.80 - 1.70    &   8.5 \%    \\
 A8-A0                 & 1.70 -  3.20   &   2.7 \%     \\
 B8-B5                 & 3.20 -  6.50   &    1.3 \%     \\\hline
\end{tabular}
\end{table}

We  checked our algorithm in predicting the close stellar encounter rate $\Gamma$ 
modeled by  Proszkow \& Adams (2009)  in their N-body simulations  of stars
in an  embedded cluster. Their power law $\Gamma = \Gamma_0 \times (b/1000 AU)^{\gamma}$  corresponds to the number
of close stellar encounters of miss distances  $< b$  per star per Myr. With our model,
we estimated the time $T_{1000AU}$ required for the first stellar flyby  at $d < 1000 AU$ 
to occur by making  $\phi~(d<1000AU) =\int_{~0}^{~T_{1000AU}}~1/t_{enc}~dt=1$ with the constant stellar 
density $n=1000$~stars~$pc^{-3}$  in eq~(2). In a similar way, we  also estimated $T_{500AU}$ 
for $d < 500$AU. With these two values  $T_{1000AU}$  and  $T_{500AU}$, we determined 
$\Gamma_0 =$ 0.4 flybys ($b < 1000$~AU) per star per Myr and $\gamma=1.5$ if $\sigma = 1$~km/s, 
and determined $\Gamma_0 =$ 1 flyby ($b < 1000$~AU) per star per Myr and $\gamma=1.9$ if $\sigma = 5$~km/s.  
These values of $\Gamma_0$ and $\gamma$ are consistent with those in Proszkow \& Adams (2009). 
As an additional test of our algorithm, we also predicted for the Sun that a close stellar encounter 
 of 0.5 pc occurs once every  1.2~Myr
in the present-day star field  ($n$=constant=0.1$~\rm star~pc^{-3}$ in eq~(2)). 
This prediction is indeed the frequency determined by  Garcia-Sanchez et al. (2001).

\begin{table}[h]
\centering
\caption{Closest stellar encounters of the Sun 
and corresponding fractions of stripped planetesimals $^{(*)}$ \label{sun}}
\begin{tabular}{l|c|c|c} 
\hline\hline\\
 Spectral type $m_p$  & $m_p/m_c$   &   $d_{enc}$    &   Fraction $s_i$ \\
                      &             &      (AU)      &                 \\\hline
                      &             &                &              \\
 M8-M5                &   0.15      &   180          &     0     \\
 M4-M0                &   0.25      &   200          &     2\%     \\
 K8-K0                &   0.5       &   230          &     3\%    \\
 G8-G0-F0             &   1.0       &   500          &     0    \\
 A8-A0                &   2.0       &   875          &     0       \\
 B8-B5                &   4.        &   1250         &     0     \\\hline
\end{tabular}

$^{(*)}$ duration=100 Myrs; $m_c= 1 \rm M_{\odot}$; $n_0=3000 \rm pc^{-3}$;  $\sigma=5$ km/s; $\phi (d <d_{enc})  =6$.

\end{table}

\subsection{Depletion of debris disks}

We turn now to our inital problem and apply Eqs.~(2), (3), and (4) to estimate the fraction of planetesimals left 
in a debris disk after close stellar encounters have occurred for the first 100~Myrs of its lifetime. 
For this estimate, we need to determine the miss distance for which six stellar flybys occur
during this period,  hence  eq (4) becomes
$\phi~(d<d_{enc})=\int_{~0}^{~t_{cl}}~1/t_{enc}~dt~=~6$ with  $t_{cl}=100$~Myr. We demonstrate below why 
six flybys are required. We solve numerically this equation 
for $d_{enc}$  and use its value in Figs \ref{strip40_100_dots} or \ref{strip10_40_dots} to estimate 
the fraction of stripped planetesimals.  This fraction is a lower limit, {\it i.e.} more planetesimals 
are ejected in reality, for two reasons.  
First, the miss distance of an encounter tracked by eq~(4) is $d_{enc}$ or smaller, and closer encounters more severely deplete a disk. 
Second, stellar flybys at 
larger miss distances that are tracked by  $\phi~(d<d_{enc})=12, 18, 24, ...$  in our algorithm are
ignored in this work, although these flybys do slightly increase the fraction of stripped planetesimals. 
This ensures that our final conclusion is conservative but not overly.
In practice, we solve numerically the equation $\phi~(d<d_{enc})=6$ above for each stellar spectral type by using
the corresponding fraction $f_p$ of Table \ref{mass_distr} that set the star number density $n$ in eq.~(3) 
and hence in $1/t_{enc}$. The fractions $f_p$ account for the fact that 
 massive passing stars are rare, but destructive for a disk, 
 and low-mass passing stars are common, but less harmful.     

We return to the choice $\phi~(d<d_{enc})=6$, which is based on     
the efficiency with which flybys strip a debris disk depending on the inclination $i$ of the passing star orbit 
 as well as the sense of circulation as discussed
in Sect.~2. As already pointed out, a stellar flyby  strips a debris disk according to the fractions
 given in Figures~\ref{strip40_100_dots} or \ref{strip10_40_dots} only if the passing star and disk motions are prograde 
and if $i < 45^{\circ}$ (two conditions). 
Assuming that the angular momentum vector of the passing star is randomly distributed over the whole celestial sphere,
the differential probability that this vector goes through a celestial band between inclinations $i$
and $i+di$ is  $dp = {1 \over 2} sin(i) di$. 
By integrating between  $i=0$ and $\pi/4$, the probability corresponding to the two conditions 
above is  $p \sim 1/6$. Hence, six flybys must occur to have one that is efficient at stripping the
planetesimals of a disk according to the fractions given in Figures~~\ref{strip40_100_dots} or \ref{strip10_40_dots}. 
 
Finally, we combine  the stripping effects over the six stellar spectral type 
ranges defined in Table \ref{mass_distr} by computing the total fraction of the planetesimals left in the disk 
after 100~Myr,  $(1-s_1) \times (1-s_2) \times .... \times (1-s_6)$, 
where each $s_i$ is the fraction of stripped planetesimals of Figures \ref{strip40_100_dots} 
or \ref{strip10_40_dots} for the appropriate ratio $m_p/m_c$.  

As an example of this calculation, we list in
Table \ref{sun} the miss distance  $d_{enc}$ and stripping factor $s_i$ for each spectral type  of the stellar encounters undergone  
by the Sun for the first 100~Myr after its birth in a cluster  characterized by $\sigma=5~km/s$ and
the initial stellar density $n_0=3000~pc^{-3}$ (Adams \& Laughlin, 2001).
In these conditions, the total fraction of planetesimals  left 
is $< 95 \%$  after 100 Myrs in our model. The present day Kuiper Belt 
has  only $\sim$ 1\% of the mass that the minimun mass solar nebula hypothesis predicts, and, the hypothesis of the late heavy
bombardement at 700~Myr has been put forth as an explaination (Morbidelli et al 2005).

In Table~\ref{strip}, we search the parameter space ($m_c$, $n_0$) 
to find out the  conditions in which a debris disk can be most severely 
depleted  during the first 100~Myr of its lifetime.
The fractions of planetesimals left after 100~Myr in a standard size disk ($40-100$~AU)  
and compact disk ($10-40$~AU) are  estimated for 
the  central star masses, $m_c= 0.25$,  0.5, 1.0 or 2.5~$M_{\odot}$, which cover 
the mass range of stars searched for debris disks 
in surveys. The cluster dispersion velocity adopted is 
$\sigma= 5$~km/s, as observed
in 11 nearby open clusters and associations with ages between 5Myrs (Upper Sco) and 
757 Myrs (Praesepe) by Madsen, Dravins \& Lindegren (2002), and as measured in
the N-body simulation of the dynamical evolution of embedded clusters from which open clusters 
emerge  (Proszkow \& Adams 2009). In Table~\ref{strip}, 
the initial star number density $n_0$ is chosen to be 100, 1000, 3000, 10~000, 20~000, or 30~000 $pc^{-3}$ 
for several reasons; most embedded clusters have
stellar densities of $\sim 100~pc^{-3}$  (Carpenter 2000, Porras et al 2003, 
Lada \& Lada 2003),  the Sun is
thought to be born in a cluster of stellar density $\sim 3000~pc^{-3}$ (Adams \& Laughlin 2001),
and  the densest and closest embedded cluster is the Orion nebula cluster with
$\sim 20~000~pc^{-3}$ (Hillenbrand \& Hartmann, 1998).  We note also that the central stellar density 
of the  Arches cluster close to the Galactic center reaches  $\sim 10^5~pc^{-3}$ (Portegies Zwart et al. 2007) but
we have not included this extreme case in our study.


\section{Discussion}

\begin{table}
\center
\caption{Fractions of planetesimals left in a debris disk after undergoing close stellar
encounters for 100~Myr in expanding open clusters of various initial star  densities.
\label{strip}}
\begin{tabular}{c|c|c|c} 
\hline\hline
     Central star          &      Cluster$^{(*)}$         & \multicolumn{2}{c}{Fraction of planetesimals left}  \\
     $m_c$  ($M_{\odot}$)  &       $n_0$  $(pc^{-3})$     &  \multicolumn{2}{c} {-------------------------------------------}  \\
                           &                              &    Disk $40-100$ AU    &      Disk $10-40$ AU       \\
                           &                              &     (\%)             &        (\%)             \\ 
\hline
                           &                              &                       &              \\
        0.25               &          100                &           $100 $       &   100       \\
        0.5                &           ``                 &          $100$        &   100       \\
        1.0                &           ``                 &          $100 $       &   100       \\
        2.5                &           ``                 &          $100$        &   100       \\
                           &                              &                       &             \\     
        0.25               &          1000                &          $100 $       &   100       \\
        0.5                &           ``                 &          $100$        &   100       \\
        1.0                &           ``                 &          $100 $       &   100       \\
        2.5                &           ``                 &          $100$        &   100       \\
                           &                              &                       &             \\    
        0.25               &          3000                &          $<65 $       &   100       \\
        0.5                &           ``                 &          $<83$        &   100       \\
        1.0                &           ``                 &          $<95$       &   100       \\
        2.5                &           ``                 &          $100$        &   100       \\
                           &                              &                       &             \\
        0.25               &         10~000               &          $< 23 $      &   100       \\
        0.5                &           ``                 &          $< 49$       &   100       \\
        1.0                &           ``                 &          $< 75$       &   100       \\
        2.5                &           ``                 &          $< 94$       &   100       \\
                           &                              &                       &             \\
        0.25               &         20~000               &          $< 3 $       &   $ < 97$   \\
        0.5                &           ``                 &          $< 15$       &   100       \\
        1.0                &           ``                 &          $< 37$       &   100       \\
        2.5                &           ``                 &          $< 58$       &   100       \\
                           &                              &                       &             \\
        0.25               &         30~000               &          $<1    $     &   $< 76 $   \\
        0.5                &           ``                 &          $<3    $     &   $<95$     \\
        1.0                &           ``                 &          $<16    $    &   100       \\
        2.5                &           ``                 &          $<40 $       &   100       \\\hline
\end{tabular}

{~~~~~~~~~~~$^{(*)}$ : $\sigma=5$ km/s, ~~$\phi (d <d_{enc})  =6$. \hfill}

\end{table}

\begin{figure}[t]
\centering
\includegraphics[width=6.0cm, angle=-90]{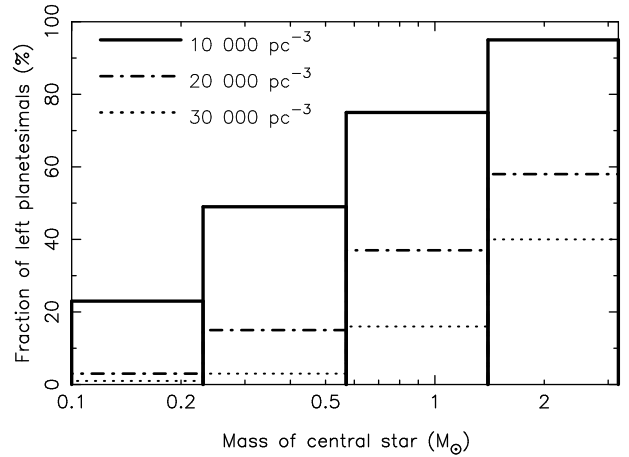}
\caption{Fractions of planetesimals left in a debris disk after undergoing close stellar
encounters for 100~Myr
in open clusters of various initial star number densities. Fractions are the same 
as in Table \ref{strip} for the disk with inner and outer radii of 40 and 100 AU.
The four bins correspond to the stellar spectral types M8-M5,
M4-M0, KGF, and A.
\label{distrib}}
\end{figure}

According to our search of the parameter space ($m_c, n_0$) in  Table \ref{strip}, severe depletion  by 
close stellar encounters occurs only for  a disk of standard size ($40-100$ AU) surrounding a star born in an embedded 
cluster with a high star-density $n_0$ greater than  $20~000~pc^{-3}$. 
In these conditions,  fewer than $58\%$ of the planetesimals are left around an intermediate-mass star after 100~Myr,  
fewer than $37\%$  around  a solar-mass star, and fewer than 
only 3\% around a low-mass star. 
In common low star-density embedded clusters  where $n_0$ is $ < 1000~pc^{-3}$,
stellar flybys have a relatively small effect on disks.
The turning point in Table~\ref{strip} where  disks start to lose their planetesimals in 100~Myr is 
the intermediate star density $n_0 \sim  3000~pc^{-3}$, which is  thought to have prevailed in the birthplace 
of the Sun.  
In contrast, disks of compact size ($10-40$~AU) are almost insensitive to their stellar environment  as seen 
in  Table \ref{strip}.
These conclusions remain qualitatively the same regardless of whether the disk is dynamically excited at the start of the simulation or not.
In agreement  with these results,  Spurzem et al. (2009) demonstrated that disruptions of some wide-orbit planetary systems  
in an Orion-type cluster are expected on a timescale of a few $10^8$ yrs, leaving free-floating planets as relics.

We can make a rough estimate of the fraction of stars  born in low and high star-density  environments 
by comparing the numbers  of stars  $N_1$  and  $N_2$ that are, respectively, in  
the closest high star-density embedded cluster,  the  Orion Nebula Cluster at 450~pc, and in 
all low star-density embedded clusters closer than this distance; these numbers are  $N_1=2520$ and  $N_2=1324$ based 
on the  catalog of Lada \& Lada (2003).   
Thus, the corresponding fractions are  $N_2/(N_1+N_2) = 34$\% and  $N_1/(N_1+N_2) = 66$\% for stars born
in  low and high star-density embedded clusters, respectively.
If this rough estimate holds over the whole Milky Way, we can conclude that about two-thirds 
of the stars surveyed today originate from high star-density embedded clusters, 
thus their disks may have suffered severe stripping    
by close stellar encounters according to Table~\ref{strip}.
We note that embedded clusters with  star-densities  higher than $10~000~M_{\odot} pc^{-3}$  
survive the initial gas expulsion phase (``infant mortality phase''), and are stable against 
disruption by stellar evolution and encounters with giant molecular clouds over an Hubble time 
(Lamers et al, 2005,  Gieles 2009).

As depicted in Figure \ref{distrib}, or equivalently in Table~\ref{strip},  the fraction of planetesimals 
left in a standard disk ($40-100$~AU)  depends significantly on the mass of the central star $m_c$. 
This is expected in our model in which flybys are more disruptive for disks around 
M-dwarfs encountering mostly more massive stars, while they are less damaging for disks around
A-dwarfs  encountering mostly less massive stars.    
Consequently, in our model, disks surrounding  M-dwarfs are expected
to be more difficult to detect since more depleted disks produce less observable dust.  
This trend in our model agrees with the observed decline in the number of 
debris disks found around lower mass stars in surveys
(32\% for A-dwarfs, 16\% for FGK dwarfs, $<5\%$ for M-dwarfs) as emphasized in Lestrade et al (2009).

A limitation of our model is that the star number density is assumed to be uniform across the cluster.
Mass segregation is a well-known feature of star clusters, as can be seen in the Trapezium where 
the most massive stars are arranged in
a central, compact core (Moeckel \& Bonnell 2009). Mass segregation can be approximated  in our
model by  a relative increase in the  high-mass star fraction $f_p$ given in Table \ref{mass_distr} 
to simulate the condition at the center of the cluster. For example,
in the case of $m_c=0.25 \rm ~M_{\odot}$ and $n_0= 20~000 \rm ~pc^{-3}$, after increasing  $f_p$ by a factor of four
for the most massive stars of this study ($m_p=2.5 \rm M_{\odot}$) and decreasing $f_p$ accordingly
for the lowest mass stars, the fraction of remaining planetesimals changes from  $<3\%$ in Table \ref{strip} to  $<2\%$. 
For the same modifications of $f_p$ but for $m_c=2.5 \rm ~M_{\odot}$ and $n_0= 20~000 \rm ~pc^{-3}$, the
fraction changes from $<58\%$ in Table \ref{strip} to  $<45\%$. 
Consequently, this limitation of our model makes our estimate conservative and leaves unchanged our main conclusion.

\section{Conclusion }

We have studied the depletion of planetesimals in a debris disk  triggered by close stellar encounters in the environment of 
an expanding open cluster over its lifetime of 100~Myrs.  
We have found that depletion is significant only for an initial star-density  of the embedded cluster of origin greater than 
$20~000~pc^{-3}$, as in the Orion Nebula Cluster, and in a disk of standard size (inner and outer radii~: 40 and 100 AU).
In these conditions, a debris disk loses $>$97\% of its  planetesimals around an M-dwarf, $>$63\%  around a solar-type star, 
and $>$42\%  around an A-dwarf in 100 Myr. 
This level of depletion  could affect two-thirds of the stars searched in  surveys since two-thirds of them 
are born in  high star-density embedded clusters  according to  the catalog of Lada and Lada (2003).
However, more compact disks (inner and outer radii~: 10 and 40~AU) are much less sensitive to their environment. 
Unfortunately, debris disk sizes are not yet well enough known  to decisively conclude whether the observed trend 
of fewer debris disks being detected around lower mass stars can be  explained by the mechanism we have studied.

\begin{acknowledgements}
We are grateful to our referee, Herv\'e Beust, for his suggestions and are indebted to him for indicating the correct probability calculation 
of the number of flybys. Etienne Morey PhD work is funded by a Fondation CFM-JP Aguilar grant.  
\end{acknowledgements}

\null
\null

\noindent References 

\bibl
Abt, H, 2009, PASP, 121, 1307 
\filbreak

\bibl
 Adams, F.C. \& Laughlin, G., 2001, Icarus, 150, 151-162
\filbreak

\bibl
Andrews, S., Wilner, D., Espaillat, C., et al, 2011, ApJ, 732, 42
\filbreak

\bibl
Augereau, J.-C., Nelson, R.P., Lagrange, A.-M., Papaloizou, J.C.B., Mouillet, D., 2001, A \& A, 370, 447
\filbreak

\bibl
Augereau, J.-C. \& Beust, H.,   2006, A \& A, 455, 987
\filbreak

\bibl
Beust, H. \& Dutrey, A., 2006, A\&A, 446, 137
\filbreak

\bibl
Binney, J. \&  Tremaine, S., 1987, Galactic Dynamics, Princeton Unversity Press
\filbreak

\bibl 
Bryden, G. C., et al., 2006, ApJ, 636, 1098
\filbreak

\bibl 
Bonnell, I.A., Kester, W.S., Melvyn, B.D., Keith, H., 2001, MNRAS, 322, 859 
\filbreak

\bibl 
Booth, M., Wyatt, M.C., Mobidelli, A., Moro-Martin, A., Levison, H.F., 2010,
MNRAS, 399, 385
\filbreak

\bibl
Carpenter, J., 2000, AJ, 120, 3139
\filbreak

\bibl
Corder, S., et al, 2009, ApJ, 690, L65-L68
\filbreak

\bibl
Converse, J.M., Stahler, S.W., 2010, MNRAS, 405, 666
\filbreak

\bibl
Dent, W. R. F., Walker,  H. J. ,  Holland, W.S.,  \& Greaves, J.S., 2000,
MNRAS., 314, 702
\filbreak

\bibl
Dominik, C. \&  Decin, G.,  2003, ApJ, 583, 626.
\filbreak

\bibl
Eiroa, C., Fedele, D., Maldonato, J., Gonzalez-Garcia, B.M., Rodmann, J., et al., 2010, A\&A, 518, L131
\filbreak

\bibl
Garcia-Sanchez, J.,  Weissman, P.R., Preston, R.A.,  Jones, D.L., Lestrade, J-F,  Latham,  D.W.,
Stefanik, R.P. and Paredes, J.M., 2001, A\&A, 379, 634
\filbreak

\bibl
Gautier, T. N.,  Rieke, G. H., Stansberry, J., Bryden, G. C., Stapelfeldt, K. R., Werner, M. W., Beichman, C. A. et al., 2007, ApJ, 667, 527
\filbreak

\bibl
Gieles, M., 2009, in {\it Star Clusters}, Proceedings of the IAU Symposium 266,  R. de Grijs \& J.R.D. L\'epine, eds, Cambridge University Press.
\filbreak

\bibl
Gomes, R., Levison, H.F., Tsiganis, K., Morbidelli, A., 2005, Nature, 435, 466.
\filbreak

\bibl
Greaves, J. S., Holland, W. S., Wyatt, M. C., Dent, W. R. F., et al, 2005, ApJ, 619, 187
\filbreak

\bibl
Hillenbrand, L. \&  Hartmann, L.W., 1998, ApJ., 492, 540 
\filbreak

\bibl
Kaib, N.A., \& Quinn, T., 2008, Icarus, 197, 221 
\filbreak

\bibl
Kalas, P., Larwood, J., Smith, B.A., \& Schultz, A., 2000, ApJ, 530, L133.
\filbreak

\bibl
Kennedy, G.M. \& Wyatt, M.C., 2010, MNRAS, 405, 1253
\filbreak

\bibl
Kenyon, S.J. \&  Bromley, B.C., 2002, ApJ, 577, L35
\filbreak

\bibl
Kenyon, S.J. \& Bromley, B.C., 2004, Nat., 432, 598
\filbreak

\bibl
Kobayashi, H. \& Ida, S., 2001, Icarus, 153, 416
\filbreak

\bibl
Lada, C.J.  \& Lada, E.A. , 2003, ARA \& A, 41, 57 
\filbreak

\bibl
Lamers, H. J. G. L. M. et al, 2005, A\&A, 441, 117 
\filbreak

\bibl
Larwood, J.D., 1997, MNRAS, 290, 490
\filbreak

\bibl
Larwood, J.D. \& Kalas, P.G., 2001, MNRAS, 323, 402-416.
\filbreak

\bibl
Lestrade, J.-F., Wyatt, M. C., Bertoldi, F., Dent, W. R. F., Menten, K. M., 2006, A\&A, 460, 733 
\filbreak

\bibl
Lestrade, J.-F., Wyatt, M. C., Bertoldi, F., Menten, K. M., Labaigt, G., 2009, A\&A, 506, 1455
\filbreak

\bibl
Liseau, R. et al., 2008, A\&A, 480, L47
\filbreak

\bibl
Madsen, S., Dravins, D., Lindegren, L., 2002, A\&A, 381, 446
\filbreak

\bibl
Malmberg, D., et al., 2007, MNRAS, 378, 1207
\filbreak

\bibl
Marshall, J.P. et al., 2011, A\&A, 529, 117
\filbreak

\bibl
Matthews, B.C., Sibthorpe, B., Kennedy, G., Philips, N., Churcher, L. et al., 2010, A\&A, 518, L135
\filbreak

\bibl 
Moeckel, N.  \& Bonnell, I.A., 2009, MNRAS, 400, 657
\filbreak

\bibl
Morbidelli, A., Levison, H.F., Tsiganis, K., Gomes, R.,  2005, Nature, 435, 462.
\filbreak

\bibl
Moro-Mart\'in, A. \& Malhotra, R. 2002, AJ, 124, 2305
\filbreak

\bibl
Mouillet, D., Larwood,J.D., Papaloizou, J.C.B, Lagrange, A.-M., 1997, MNRAS, 292, 896
\filbreak

\bibl
Mustill, A.  \& Wyatt, M.C., 2009, MNRAS, 399, 1403-1414
\filbreak

\bibl
Najita, J. \&  Williams, J.P., 2005, ApJ, 635, 625 
\filbreak

\bibl
Olczack, C., Pfalzner, S. \& Eckart, A.,  2009, A\&A, 509, 630
\filbreak

\bibl
Porras, A. et al, 2003, AJ, 126, 1916 
\filbreak

\bibl
Portegies Zwart, S., et al., 2007, MNRAS, 378, L29.
\filbreak

\bibl
Press, W.H., Teukolsky, S.A., Vetterling, W.T., Flannery, B.P., 1992, Cambridge University Press. 
\filbreak

\bibl
Proszkow, E.-M., Adams, F.C., 2009, ApJS, 185, 486.  
\filbreak

\bibl
Reche, R., Beust, H., Augereau, J.-C., 2009, A \&A, 493, 661 
\filbreak

\bibl
Spurzem, R, Giersz, M, Heggie, D. C., Lin, D.N.C., 2009, ApJ, 697, 458 
\filbreak

\bibl
Su, K.Y.L.,  Rieke, G. H.,  Stapelfeldt, K.R., Stansberry, J.A.,  Bryden, G., Stapelfeldt, K.R., et al.  2006, ApJ, 653, 675
\filbreak

\bibl
Toomre, A.  \& Toomre, J., 1972, ApJ, 178, 623  
\filbreak

\bibl
Trilling, D. E., Bryden, G., Beichman, C. A., Rieke, G. H., Su, K. Y. L., Stansberry, J. A., et al., 2008, ApJ, 674, 1086 
\filbreak

\bibl
Tsiganis, K., Gomes, R.,  Morbidelli, A., Levison, H.F., 2005, Nature, 435, 459.
\filbreak

\bibl
Wyatt, M.C., 2003, ApJ, 598, 1321
\filbreak

\bibl
Wyatt, M.C. \&  Dent, W.F.R.,  \& Greaves, J.S.,  2003, MNRAS, 342, 876.
\filbreak

\bibl
Wyatt, M.C., 2008, ARA\&A, 46, 339-383
\filbreak

\end{document}